\documentclass[12pt,aps,prb,preprint]{revtex4}   

\usepackage{amsmath}    
\usepackage{amssymb}
\usepackage{graphicx}   

\usepackage[latin1]{inputenc}     
\usepackage[T1]{fontenc}

\usepackage{float}   

\draft

\begin{document}

\title{Energy flow lines and the spot of Poisson-Arago}
\author{Michel Gondran}
\email{michel.gondran@polytechnique.org} \affiliation{University
Paris Dauphine, Lamsade, 75 016 Paris, France}
\author{Alexandre Gondran}
 \affiliation{SET, Université Technologique de Belfort Montbéliard}

\begin{abstract}
We show how energy flow lines answer the question about
diffraction phenomena presented in 1818 by the French Academy:
"\textit{deduce by mathematical induction, the movements of the
rays during their crossing near the bodies"}. This provides a
complementary answer to Fresnel's wave theory of light. A
numerical simulation of these energy flow lines proves that they
can reach the bright spot of Poisson-Arago in the shadow center of
a circular opaque disc. For a monochromatic wave in vacuum, these
energy flow lines correspond to the diffracted rays of Newton's
\textit{Opticks}.
\end{abstract}

 \maketitle

\section{Introduction} \label{}

\bigskip

The answer Fresnel provided in 1818 in response to the French
Academy's competition marks the beginning of the refutation of
Newton's corpuscular theory of light and the rehabilitation of the
Huygens wave theory. The competition topic was presented as
follows:

"\textit{...diffraction phenomena have been a subject of research
for many physicists...but research has not yet sufficiently
determined the movement of the rays near the body where the change
occurs...it important...to further study...the physical manner in
which rays are inflected and separated into different bands ... As
a result the Academy is proposing this research...to be presented
as follows: 1. Determine all the effects of ray diffraction
...direct and reflected when they ... pass near the extremities of
a body... 2. Deduce from these experiments, by mathematical
induction, the movements of the rays during their crossing near
the bodies.}"~\cite{Fresnel_1866}

This announcement was made by a jury of great scientists:
Pierre-Simon Laplace, Jean B. Biot, Simeon D. Poisson, Joseph L.
Gay-Lussac - all Newtonian - as well as Dominique F. Arago who was
the only one who believed in wave theory.
\begin{figure}[H]
\begin{center}
\includegraphics[width=0.4\linewidth]{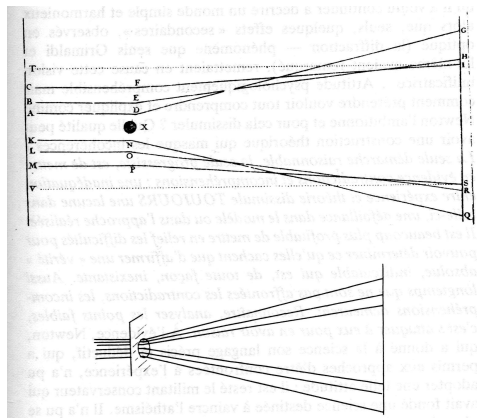}
\caption{\label{fig:schemexperience} Newton's rays diffracted by a
hair and a small circular aperture (1704).~\cite{Newton_1704}}
\end{center}
\end{figure}
They all recall the ray concept proposed by Newton in the third
book of his \textit{Opticks} ~\cite{Newton_1704} [see Fig.
\ref{fig:schemexperience}] in order to explain diffraction by a
hair or by a circular aperture, and the conclusion of its
experimental part where he wrote:

"\textit{When I made the foregoing Observations, I design'd to
repeat most of them with more care and exactness, and to make some
new ones for determining the manner how the Rays of Light are bent
in their passage by Bodies, for making the Fringes of Colours with
the dark lines between them. But I was then interrupted, and
cannot now think of taking these things into farther
Consideration. And since I have not finish'd this part of my
Design, I shall conclude with proposing only some Queries, in
order to a further search to be made by
others.}"~\cite{Newton_1704}

Fresnel's essay develops a mathematical wave theory which seems be
in conflict with the corpuscular theory. It describes an
impressive number of diffraction experiments all explained by the
same principle: the fringes are due to interference waves issued
by each of the screen points. Fresnel's principle generalizes
Huygens' principle.

Poissson carefully studied Fresnel's theory and deduced
"\textit{that the center of the shadow of a circular opaque
disc... (should)... be as enlightened as if the disc didn't
exist}"~\cite{Fresnel_1866}; this bright spot of light at the
center of the shadow, he claimed, "\textit{violated common sense}"
and hence refuted Fresnel's wave theory. However, Arago almost
immediatly verified the spot experimentally. Fresnel won the
competition and this discovery induces the acceptance of wave
theory and the refutation of corpuscular theory. This spot of
light, today known as \textit{ Poisson's bright spot} or
\textit{spot of Arago}, was observed a century earlier (1723) by
Maraldi, who had not published his
work.~\cite{Harvey_1984,Kelly_2009}

This paper proposes to complete Fresnel's answer and to show how
the energy flow lines are (in the special case of a monochomatic
wave in vacuum) the answer to the French Academy's question about
diffraction phenomena:"\textit{deduce by mathematical induction,
the movements of the rays during their crossing near the bodies"}.
We study, by a numerical simulation, the case of diffraction by a
circular aperture as well as diffraction by a circular opaque disc
in order to find the spot of Poisson-Arago. In Section 2, we
recall how to calculate bright densities with wave theory. In
Section 3, we show that the energy flow lines correspond to Newton
rays and are in good agreement with the experiment. Then in
section 4, we discuss the interpretation of these energy flow
lines.

\section{Intensity distributions}

Let us consider a monochromatic plane wave of light which is
perpendicular to a circular aperture (resp. an opaque disc) placed
in the xy plane, and a detector in a parallel plane at distance z.
Set ($x_M,y_M$) the coordinates of a point M in the diffracting
plane and (x, y) the coordinates of the observation point P on the
detector.

First, if we neglect the polarisation of light and if we suppose
that the incident wave is of the form $A_0 e^{ikz}$ on the
aperture, the amplitude $A(P)$ for $z>0$, which verifies the
Helmholtz equation, is given by the Rayleigh-Sommerfeld
formula:~\cite{Sommerfeld_1894,Rayleigh_1897}
\begin{equation}\label{eq:eqapresouverture}
A(P)= -\frac{i A_0}{\lambda} \int_{S}  \frac{e^{ikr}}{r} (1-
\frac{1}{i k r})\cos \theta dx_M dy_M
\end{equation}
where $r= \sqrt{(x-x_M)^2 + (y-y_M)^2 +z^2}$, $\cos\theta=
\frac{z}{r}$, $k=\frac{2 \pi}{\lambda}$ and where the integration
is taken on the surface S of the aperture. Notice that the formula
gives the exact solution, in particular for very small distance of
the aperture thanks to $- 1 /i k r $; see Gillen and
Guha.~\cite{Gillen_2004}

\subsection{Intensity distributions for a circular
aperture}

\begin{figure}[H]
\begin{center}
\includegraphics[width=0.5\linewidth, height=0.2\linewidth]{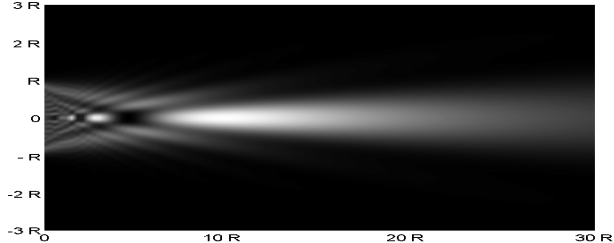}
\caption{\label{fig:diffractiontrou} Calculated intensity
distributions behind the circular aperture in the plane (z, x). }
\end{center}
\end{figure}

Figure \ref{fig:diffractiontrou} shows, in a plane (z,x)
containing the optical axis, intensity behind a circular aperture
with a radius $R=5 \mu m $ of a monochromatic plane wave of light
with wavelength $\lambda= \frac{R}{10}= 0.5 \mu m $.
Far from the aperture, the classical Fraunhofer diffraction
appears, and the emerging beam has a well-defined angular
dispersion, in the order of $\Delta\theta\sim\frac{\lambda}{R}$;
this is the Airy disc. Near the aperture, we get the Fresnel
diffraction. We note a succession of bright and dark areas on the
axis.

\subsection{Intensity distributions for an circular opaque disk}

Because the incident wave is a plane wave, the intensity value is
calculated by the Babinet's principle~\cite{Babinet_1837, Ganci
2005} in taking the square of
\begin{equation}\label{eq:eqapresdisque}
A(P) = A_0( e^{i k z} + \frac{i }{ \lambda} \int_{S}
\frac{e^{ikr}}{r} (1- \frac{1}{i k r} )\cos \theta dx_M dy_M);
\end{equation}
numerical integration is taken on the surface of the opaque disc
S.

\begin{figure}[H]
\begin{center}
\includegraphics[width=0.5\linewidth, height=0.2\linewidth]{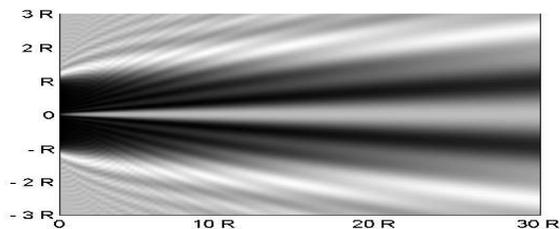}
\caption{\label{fig:diffractiondisque} Calculated intensity
distributions behind the circular opaque disk in the plane (z,
x).}
\end{center}
\end{figure}

Figure \ref{fig:diffractiondisque} represents the intensity behind
an opaque disc of radius $R=5 \mu m $ of a monochromatic wave of
light with wavelength $\lambda= \frac{R}{10}= 0.5~ \mu m $.
We see clearly the area corresponding to the geometric shade, but
also the bright spot of Poisson-Arago in the center of the shadow;
with the choice of $\lambda =\frac{R}{10}$, this bright spot
assumes a great importance.

\begin{figure}[H]
\begin{center}
\includegraphics[width=0.3\linewidth]{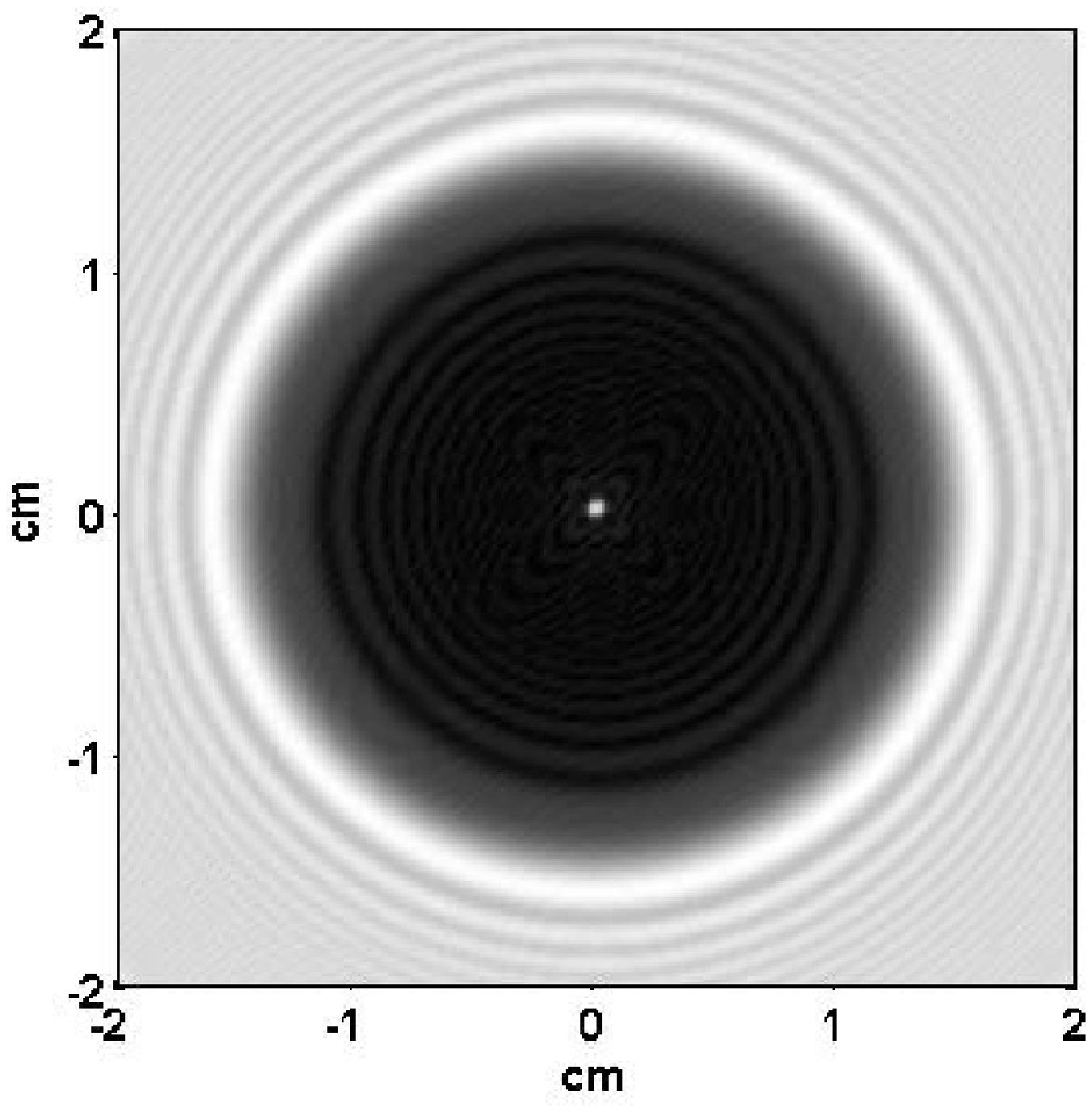}
\includegraphics[width=0.3\linewidth]{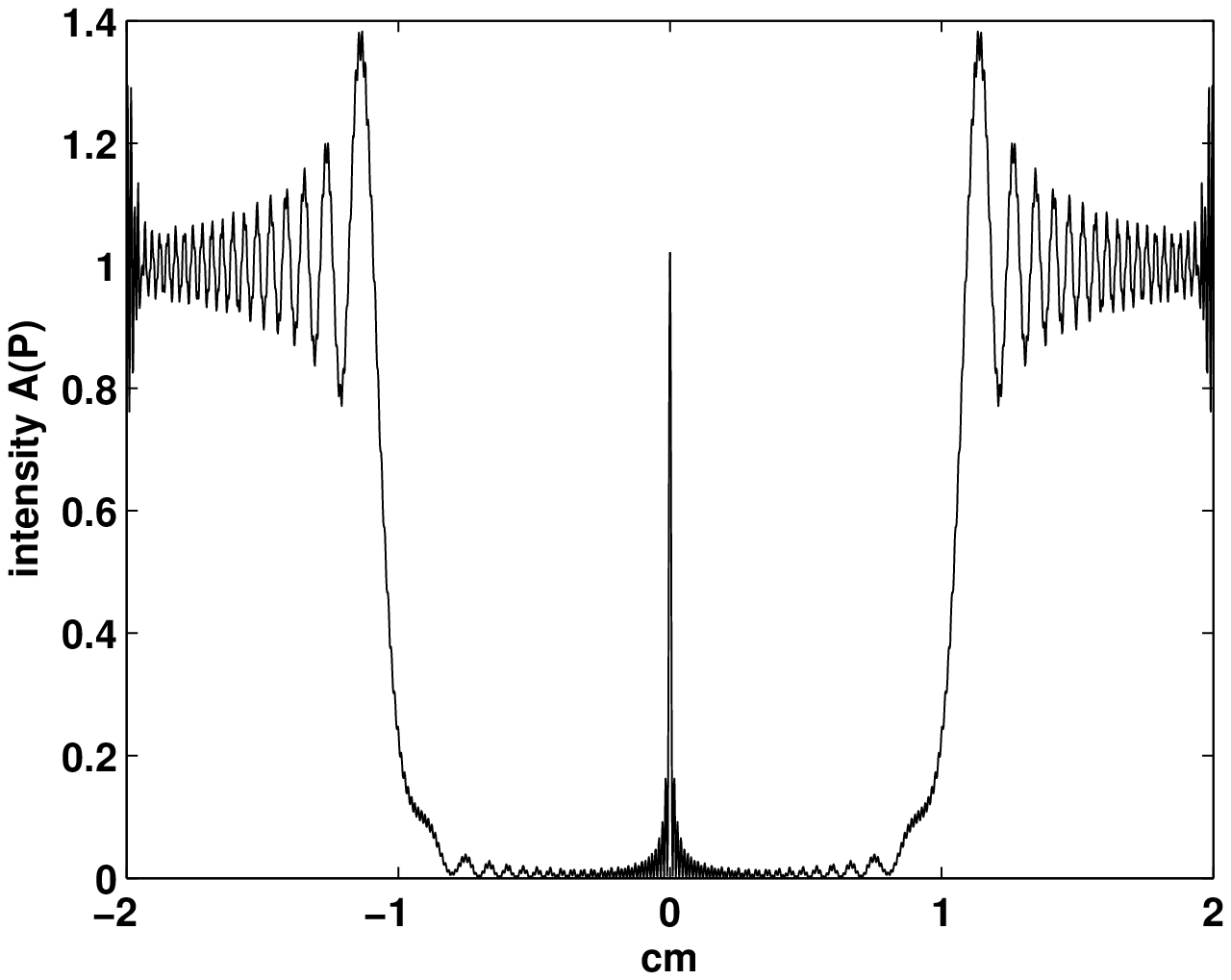}
\caption{\label{fig:diffractiondisquearago} Spot of Poisson-Arago:
Intensity distribution behind a coin ($R = 1~ cm $) on a detector
placed at 5m.}
\end{center}
\end{figure}

Newton, who carried out the experiment of the opaque disk using a
coin, does not report the presence of fringes within the shadow in
his \textit{Opticks}.~\cite{Newton_1704} With $\lambda= 0.5~ \mu m
$ and a coin ($R = 1~ cm $), it is difficult to see the bright
spot of Arago on a detector placed at 5 m [see Fig.
\ref{fig:diffractiondisquearago}]. The radius of the spot is 0.1
mm and just visible to the naked eye. This oversight was to have
an unfortunate consequence a century later!

\section{Energy flow lines for a monochromatic wave}

Let us consider a monochromatic electromagnetic field $
\{\textbf{E}(\textbf{r},t),\textbf{ B}(\textbf{r},t) \}$ which is
the real part of the complex monochromatic electromagnetic field $
\{ \mathcal{E}(\textbf{r},t)= \textbf{E}_0(\textbf{r}) e^{- i
\omega t}, \mathcal{B}(\textbf{r},t)= \textbf{B}_0(\textbf{r})
e^{- i \omega t}\}$.

The Poynting vector $\textbf{S}= \frac{1}{\mu_0} \textbf{E} \times
\textbf{B}$ is the instantaneous rate of energy flow per unit area
at a point; $u=\frac{1}{2}( \epsilon_0 \textbf{E}^2 +
\frac{1}{\mu_0} \textbf{B}^2)$ is the instantaneous
electromagnetic energy density. Since the optical frequencies are
very large ($ \omega$ is of order $10^{15} $ $s^{-1}$), one cannot
observe the instantaneous values of any of the rapidly oscillating
quantities, but only their time average taken over a time interval
which is large compared to the fundamental period
$T=\frac{2\pi}{\omega}$.~\cite{Born, Jackson_1999}

The energy flow, which is interpreted as the time-averaged
Poynting vector,~\cite{Born, Jackson_1999} is determined from the
real part of the complex Poynting vector $
\mathcal{S}=\frac{1}{\mu_0}\mathcal{E} \times \mathcal{B}^{*}$ and
from the energy density of the complex field $\mathcal{U} =
\frac{1}{2} ( \varepsilon_0 \mathcal{E}\mathcal{E}^{*}+
\frac{1}{\mu_0} \mathcal{B}\mathcal{B}^{*}) $. The time-averaged
flux of energy and the time-averaged energy density are given by
\begin{equation}\label{eq:eqchampdensite}
\langle \textbf{S}\rangle=\frac{1}{2 \mu_0} Re [\textbf{E}_0
\times \textbf{B}_0^{*}],~~~~\langle u \rangle=\frac{1}{4} (
\varepsilon_0 \textbf{E}_0\textbf{E}_0^{*}+ \frac{1}{\mu_0}
\textbf{B}_0 \textbf{B}_0^{*}).
\end{equation}
The energy flow lines are obtained by the equation
\begin{equation}\label{eq:eqvitesse}
\frac{d \textbf{r}}{dt}= \frac{\langle \textbf{S}\rangle}{\langle
u \rangle}.
\end{equation}

For diffraction problems, these energy flow lines have been
discussed at length. In 1952, they were calculated numerically for
two-dimensional diffraction on a half-plane by Braunbeck and
Laukien~\cite{Braunbeck_1954} and recalled in Born and Wolf's
textbook~\cite{Born} p.575-577. In 1976,
Prosser~\cite{Prosser_1976} proposed an interpretation of
diffraction and interference with electromagnetic fields in terms
of energy flow lines. These lines are recently demonstrated and
discussed in distributions of incoherent light for various
two-dimensional situations by Wünscher et al.~\cite{Wunschen_2002}
Their interpretation will be discussed in the following section.

The circular aperture and opaque disc problems are invariant by
rotation, and if the incident light polarization is also invariant
by rotation, the electromagnetic field ($\textbf{E}_0,
\textbf{B}_0$) can be written in cylindrical coordinates
($\rho,\varphi,z$) in the form $\textbf{E}_0= \{e_\rho,0,e_z\}$
and $\textbf{B}_0= \{0,b_\varphi,0\}$, where $e_\rho$, $ e_z$ and
$b_\varphi$ are functions of ($\rho,z$).

From Maxwell equation $rot \mathcal{B}= \varepsilon_0 \mu_0
\frac{\partial \mathcal{E}}{\partial t}$ we deduce $\textbf{E}_0=
\frac{i c}{k}[-\frac{\partial}{\partial
z}b_{\varphi},0,\frac{\partial}{\partial\rho} b_{\varphi} ]$ and
\begin{equation}\label{eq:eqvecteurpointing}
\langle \textbf{S} \rangle= \frac{1}{2 \mu_0}\frac{c \lambda}{2
\pi} Im(b_{\varphi}^* \nabla b_{\varphi} ).
\end{equation}
From Faraday's law $rot \mathcal{E}= - \frac{\partial
\mathcal{B}}{\partial t}$, we show that $b_\varphi $ verifies the
Helmholz equation
\begin{equation}\label{eq:eqhelmholtz}
\Delta b_\varphi(\rho,z) + k^2 b_\varphi(\rho,z)=0.
\end{equation}
Taking into account the time-averaged energy conservation law
$\nabla \cdot \langle \textbf{S} \rangle =0$, we deduce
$\varepsilon_0 \textbf{E}_0\textbf{E}_0^{*}= \frac{1}{\mu_0}
\textbf{B}_0 \textbf{B}_0^{*}$ and then $\langle u \rangle=
\frac{1}{ 2 \mu_0} b_\varphi b_\varphi^{*}$.

The energy flow lines are then only defined by the wave
$b_\varphi$
\begin{equation}\label{eq:eqvitesse}
\frac{d \textbf{r}}{dt}= \frac{c \lambda}{2 \pi} \frac{Im(
b_\varphi^* \nabla b_\varphi )}{b_\varphi b_\varphi^{*}}
\end{equation}
and are perpendicular to equal phase surfaces; if $ b_\varphi=
|b_\varphi| \exp(i \theta) $, $\nabla\theta = Im( b_\varphi^*
\nabla b_\varphi )/ b_\varphi b_\varphi^{*} $.

\subsection{Energy flow lines for a circular
aperture}

\begin{figure}[H]
\begin{center}
\includegraphics[width=0.5\linewidth, height=0.2\linewidth]{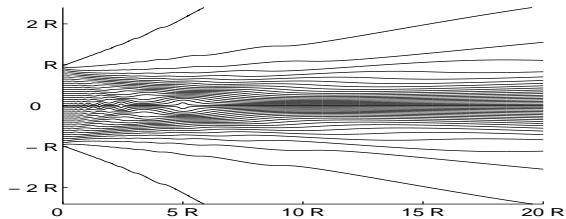}
\caption{\label{fig:trajdiffractionouverture} 40 energy flow lines
behind the circular aperture.}
\end{center}
\end{figure}

Figure \ref{fig:trajdiffractionouverture} shows 40 energy flow
lines where the initial positions are drawn at random in the
circular aperture.
\begin{figure}[H]
\begin{center}
\end{center}
\end{figure}

We notice that after a disturbance in the Fresnel zone, lines
gradually become straight in the Fraunhofer area, in agreement
with the diffracted rays proposed by Newton in Figure
\ref{fig:schemexperience} for a circular aperture.

\subsection{Spot of Poisson-Arago and energy flow lines for a circular opaque disk}

\begin{figure}[H]
\begin{center}
\includegraphics[width=0.5\linewidth, height=0.2\linewidth]{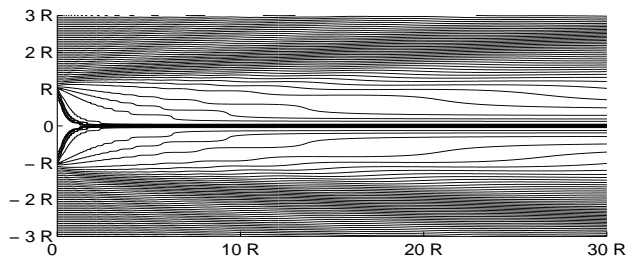}
\caption{\label{fig:trajdiffractionouverture2} Energy flow lines
behind the circular opaque disk.}
\end{center}
\end{figure}

Figure \ref{fig:trajdiffractionouverture2} shows energy flow lines
where the initial positions are drawn at random outside the
circular opaque disk.

Using the presence of energy flow lines behind the opaque disc, we
propose an explanation of the bright spots of Poisson-Arago in the
next section.

\subsection{Energy flow lines for Young's double slit experiment}

Complete these numerical simulations by determining the energy
flow lines for the Young's double slit experiment. Carried out in
1802 by Thomas Young, some years before Fresnel's theory, this
well-know experiment is the first that clearly demonstrates the
wave nature of light.\cite{Young_1802}

Let us consider a monochromatic plane wave of light ($\lambda=
0.5~ \mu m $) perpendicular to two slits placed in the xy plane,
and a detector in a parallel plan at distance z. The slits have a
width $ d= 5~ \mu m $ along x and infinity along z; 2 $d$ is the
distance between slits, center to center.

The electromagnetic field ($\textbf{E}_0, \textbf{B}_0$) is
function of $(x,z)$ and can be written $\textbf{B}_0= [0,0,
b_{z}]$, $\textbf{E}_0= \frac{i c}{k}[\frac{\partial}{\partial
y}b_{z},- \frac{\partial}{\partial x} b_{z},0 ]$. The energy flow
lines after the slits are then  defined by
\begin{equation}\label{eq:eqavitesse}
\frac{d \textbf{r}}{dt}= \frac{c \lambda}{2 \pi} \frac{Im( b_z^*
\nabla b_z )}{b_z b_z^{*}}.
\end{equation}

If the incident wave $b_z$ is of the form $A_0 e^{ikz}$ on the
slits, $b_z$ is, after the slits, given by the Fresnel-Kirchhoff
solution:
\begin{equation}\label{eq:eqapresslits}
b_z(P)= \frac{ A_0}{\sqrt{\lambda z}}e^{-i \frac{\pi}{4}}
e^{ikz}\int_{S} e^{\frac{ik (x -x_M)^2}{2 z}} dx_M
\end{equation}
where the integration is taken on the length S of the two slits.

\begin{figure}[H]
\begin{center}
\includegraphics[width=0.5\linewidth, height=0.2\linewidth]{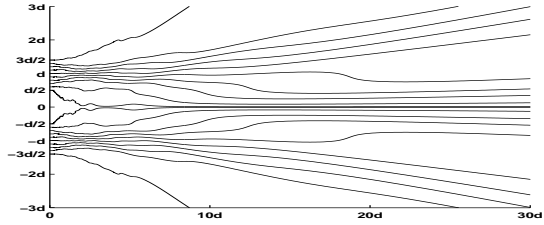}
\caption{\label{fig:trajslits} 20 energy flow lines behind the two
slits.}
\end{center}
\end{figure}

Figure \ref{fig:trajslits} shows 20 energy flow lines where the
initial position are drawn at random in the two slits.

\section{Interpretation of monochromatic energy flow lines}

For monochromatic waves in the vacuum, these energy flow lines
correspond to the diffracted rays proposed by Newton in
\textit{Principia}~\cite{Newton_1687} (1687):"\textit{Moreover,
the rays of light that are in our air (as lately was discovered by
Grimaldi, by the admission of light into a dark room through a
small hole, which I have also tried) in their passage near the
angles of bodies, whether transparent or opaque (such as the
circular and rectangular edges of gold, silver and brass coins, or
of knives, or broken pieces of stone or glass), are bent or
inflected round those bodies as if they were attracted to them.}"

Can these energy flow lines be interpreted as light rays, or even
photon trajectories?

When light is incoherent, we must reject this interpretation as
recalled by Wünscher et al.~\cite{Wunschen_2002} Indeed, when the
light is not monochromatic, it must be regarded as a mixture of
monochromatic waves as Newton showed in his famous experiments of
light decomposition.~\cite{Newton_1704} Each monochromatic wave of
white light gives energy flow lines which depend on its
wavelength.

\begin{figure}[H]
\begin{center}
\includegraphics[width=0.3\linewidth, height=0.2\linewidth]{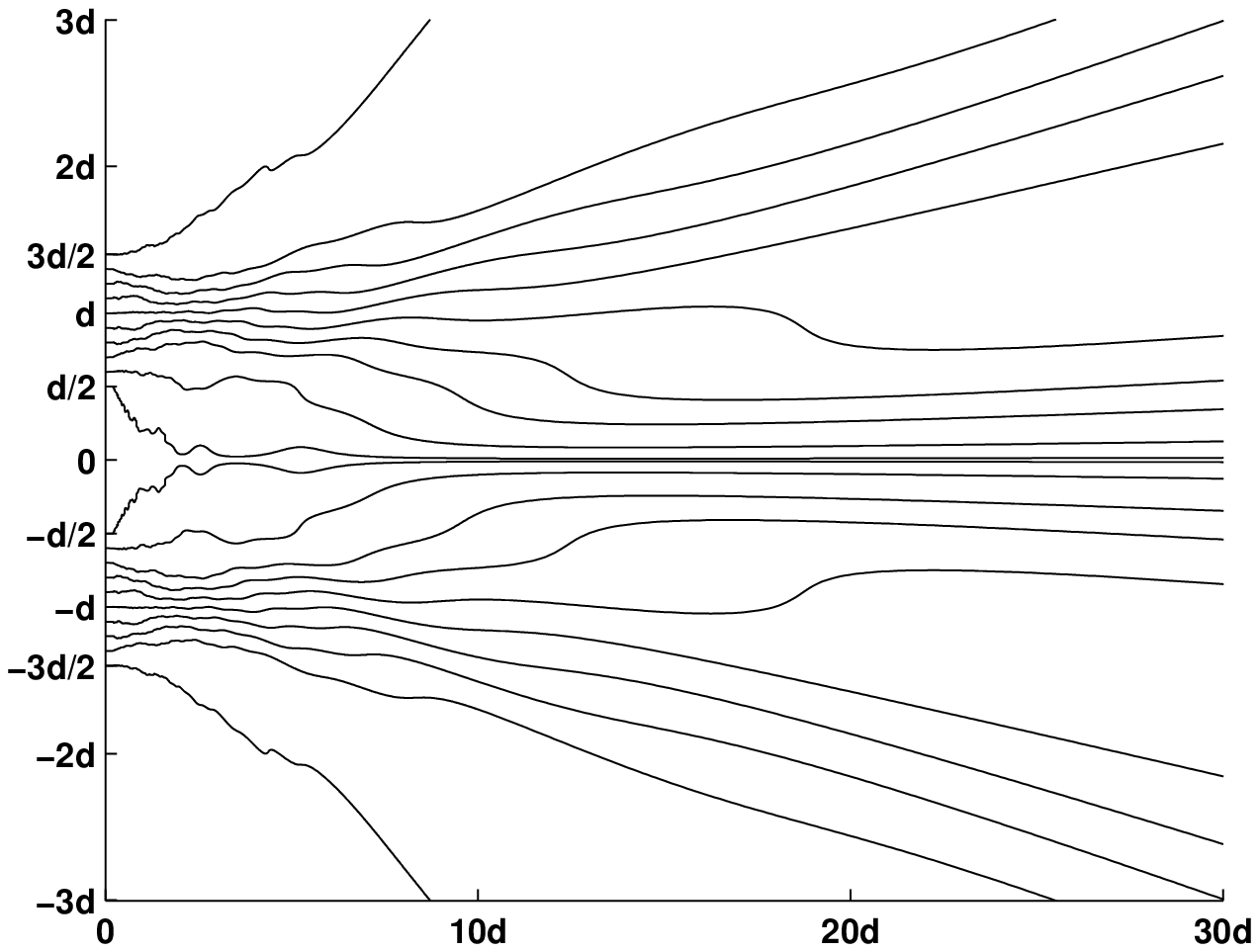}
\includegraphics[width=0.3\linewidth, height=0.2\linewidth]{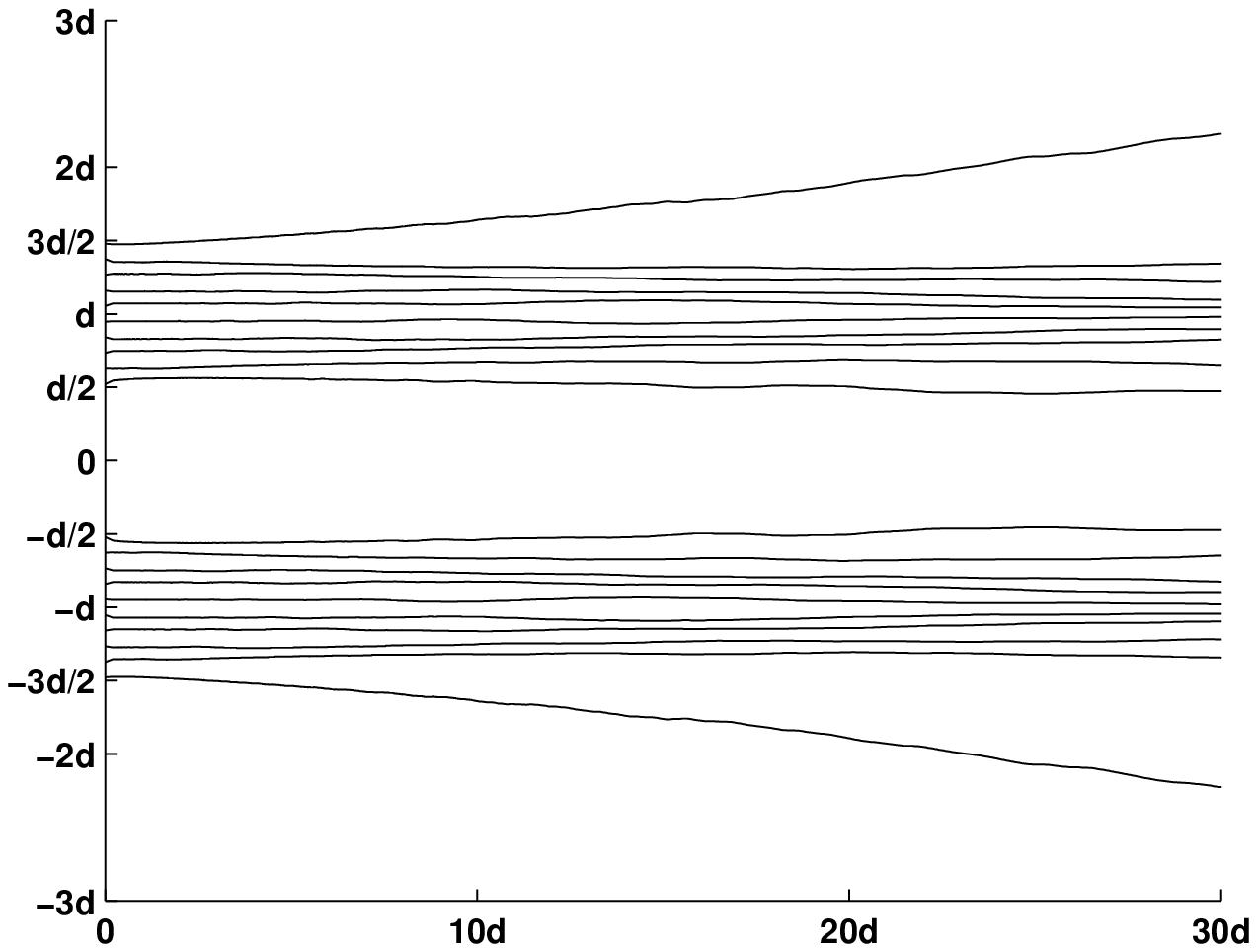}
\includegraphics[width=0.3\linewidth, height=0.2\linewidth]{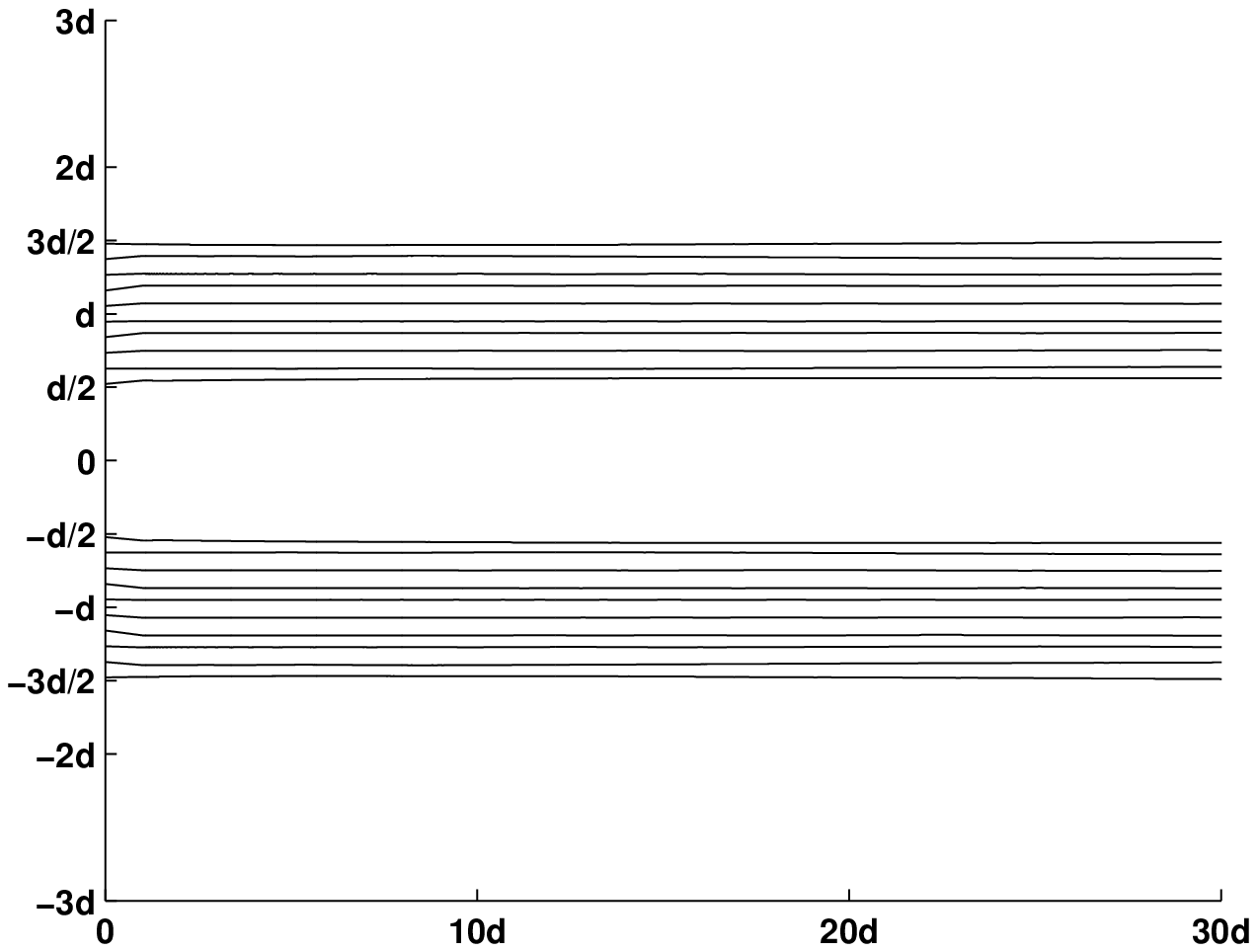}
\caption{\label{fig:convtrajslits} Evolution of the energy flow
lines when the frequency increases: $\lambda= 0.5~ \mu m $;
$\lambda= 50~ n m $; $\lambda= 5~ n m $.}
\end{center}
\end{figure}

The answer is more complex for a monochromatic wave. These energy
flow lines are a generalization of the rays of the geometrical
optics. Indeed, if we increase the frequency of the light wave
towards infinity, the energy flow lines converge towards the
straight rays of the geometrical optics. This is shown in Fig.
\ref{fig:convtrajslits} for Young's double slit interference.
Since in geometrical optics we speak of the light rays, the energy
flow lines for monochromatic waves in vacuum could be called by
analogy, the light rays of wave optics. These energy flow lines
correspond to the definition of rays of light given by Newton in
the begining of his Opticks: " \textit{By the Rays of Light I
understand its least Parts, and those as well Successive in the
same Lines as Contemporary in several Lines.}"
Could these light rays could also be interpreted as photon
trajectories? This is the question recently posed by Davidovic et
al.~\cite{Davidovic_2009}

First, these chromatic lines can be considered as photon
trajectories only if they correspond to instantaneous energy flow
lines, and not to average energy flow lines. It is possible if the
physical electromagnetic field is not the real field
$\{\textbf{E}(\textbf{r},t),\textbf{ B}(\textbf{r},t) \} $, as it
is supposed in all the classical electromagnetic
textbooks,~\cite{Born, Jackson_1999} but the complex field $ \{
\mathcal{E}(\textbf{r},t)= \textbf{E}_0(\textbf{r}) e^{- i \omega
t}, \mathcal{B}(\textbf{r},t)= \textbf{B}_0(\textbf{r}) e^{- i
\omega t}\}$.

While $Re (\mathcal{ S})/ \mathcal{U}= \langle \textbf{S}\rangle/
\langle u \rangle$, the Eq. (\ref{eq:eqvitesse}) of the average
energy flow line of the real field is also the equation of the
instantaneous energy flow line of the complex field
\begin{equation}\label{eq:eqavitessecomplexe}
\frac{d \textbf{r}}{dt}= \frac{Re (\mathcal{ S})}{\mathcal{U}}.
\end{equation}
Thus for a monochromatic light in a vacuum, the calculations show
that energy flow lines correspond to photon trajectories. Although
this hypothesis contradicts the common interpretation, we will
show that this hypothesis does not disagree with the basis of the
light quantization of quantum optics. To quantify the electric
field $\textbf{E}(\textbf{r},t)$, Glauber~\cite{Glauber_1963}
decomposes it on the one hand in monochromatic waves (it is the
Newton decomposition), on the other hand into its positive and
negative frequency parts: $\textbf{E}(\textbf{r},t)=\textbf{E}^{\{
+\}}(\textbf{r},t) + \textbf{E}^{\{ -\}}(\textbf{r},t)$. In the
monochromatic case, we have $
\textbf{E}(\textbf{r},t)=\textbf{E}_0^{\{ +\}}(\textbf{r}) e^{- i
\omega t}+ \textbf{E}_0^{\{ -\}}(\textbf{r})e^{+ i \omega t}$ with
$\textbf{E}_0^{\{ +\}}(\textbf{r})=\textbf{E}_0 (\textbf{r})$, $
\textbf{E}_0^{\{ -\}}(\textbf{r})= \textbf{E}_0^*(\textbf{r})$.
The quantization of $\textbf{E}_0^{\{ +\}}(\textbf{r})$ is then
realized with a photon annihilation operator, and the quantzation
of $\textbf{E}_0^{\{ -\}}(\textbf{r})$ with a photon creation
operator. Consequently, the electric field operator
$\widehat{\textbf{E}}(\textbf{r},t)$ is deduced by quantification
of the complex field $ \mathcal{E}(\textbf{r},t)$ (and its
conjugate), and not of the real field $\textbf{E}(\textbf{r},t)$.
It is consistent with the option of taking the complex field as
physics field. Light beams in the above examples do not correspond
to a stationary field as in the cavities, but are continually
produced by a source and, after a free propagation, continually
destroyed by absorption on the detector.

These photon trajectories, if they exist, are defined by the Eqs.
(\ref{eq:eqvitesse}), (\ref{eq:eqavitesse}) or
(\ref{eq:eqavitessecomplexe}) and by adding positions to the
monochromatic wave function. These photon trajectories are
analogous with the trajectories of massive particles of the
Broglie-Bohm
interpretation.~\cite{deBroglie,Bohm,BohmHolland,Gondran_2005}

\section{Conclusion}

The energy flow lines concept is the simplest answer to the
question of the French Academy : "\textit{deduce by mathematical
induction, the movements of the rays during their crossing near
the bodies"}. These lines correspond to the diffracted rays
proposed by Newton, and by analogy to the geometrical optics, they
can be also considered as the light rays of wave optics. So the
"spot of Poisson-Arago" could be explained by the effect of these
rays. Finally, the mathematical and numerical developments of this
paper show that Fresnel's wave theory may not be in contradiction
with the corpuscular interpretation.

\end{document}